\documentclass[preprint,prb,showpacs,amsmath,amssymb]{revtex4}

\usepackage{graphicx}% Include figure files
\usepackage{dcolumn}% Align table columns on decimal point
\usepackage{bm}% bold math
\usepackage{color}

\begin{document}

\title{Unconditionally secure one-way quantum key distribution using
decoy pulses}

\author{Z. L. Yuan}
%\email{zhiliang.yuan@crl.toshiba.co.uk}

\author {A. W. Sharpe}

\author{A. J. Shields}

\affiliation{Toshiba Research Europe Ltd, Cambridge Research
Laboratory, 260 Cambridge Science Park, Milton Road, Cambridge, CB4
0WE, UK }

\date{\today}% It is always \today, today,

\begin{abstract}
We report here a complete experimental realization of one-way
decoy-pulse quantum key distribution, demonstrating an
unconditionally secure key rate of 5.51 kbps for a 25.3~km fibre
length. This is two orders of magnitudes higher than the value that
can be obtained with a non-decoy system. We introduce also a simple
test for detecting the photon number splitting attack and highlight
that it is essential for the security of the technique to fully
characterize the source and detectors used.
\end{abstract}

% \pacs{Valid PACS appear here}% PACS, the Physics and Astronomy

\maketitle

%\section{Introduction}

Communication with single photons promises for the first time a
method to exchange digital keys between two remote parties (referred
to as Alice and Bob),\cite{bennett84,gisin02,dusek06} the secrecy of
which are not reliant upon assumptions about an eavesdropper's
surveillance equipment, computing power or
guile.\cite{mayers01,shor00} Currently, however, the security of
quantum key distribution (QKD) is compromised by the multi-photon
pulses inevitably produced by the attenuated lasers used in today's
systems.\cite{townsend93,gobby04a} In the photon number splitting
(PNS) attack, an eavesdropper (Eve) preferentially transmits these
multi-photon pulses to Bob through a lossless channel after removing
part for measurement, while blocking a correponding number of single
photon pulses to maintain the same detection rate at
Bob.\cite{brassard00} By doing so, she can determine part, or all,
of the key while remaining hidden.

Recently it has been proposed that the PNS attack may be foiled
using decoy pulses, with a different average intensity from the
signal pulses.\cite{hwang03,wang05,lo05,ma05} By determining the
transmission of the signal and decoy pulses separately, Alice and
Bob are able to detect PNS attacks.  An early
experiment\cite{zhao06} has already demonstrated the feasibility of
the decoy pulse protocol, although using a send-and-return
architecture for which Alice does not have full control of the
individual pulse intensities, as required for its secure
implementation. We report here a complete experimental demonstration
of one-way decoy pulse QKD over 25.3~km fibre with a continuous key
rate of 5.51 kbps, which is unconditionally secure against all types
of attacks, including PNS.

Decoy pulse QKD theory gives a rigorous bound of the characteristics
of the single photon pulses, which are the only source pulses that
contribute to the secure bit rate.\cite{inamori01, gottesman04} In
the simplest decoy protocol,\cite{ma05} two different average pulse
intensities (referred to as signal and decoy pulses) with mean
photon numbers of $\mu$ and $\nu$ ($\mu>\nu$) respectively are used.
By measuring the transmittances $Q_\mu$ and $Q_\nu$, \textit{i.e.},
Bob's detection probability of a signal or decoy pulse respectively,
the lower bound ($Q_1^L$) of the single photon transmittance, which
is defined as the joint probability that a signal pulse contains
only one photon and the pulse is detected by Bob, can be written
as\cite{ma05,zhao06}
\begin{equation}
Q_1^L=\frac{\mu^2e^{-\mu}}{\mu\nu-\nu^2}(Q_\nu^Le^\nu-Q_\nu
e^{\mu}\frac{\nu^2}{\mu^2}-\varepsilon_\mu Q_\mu
e^{\mu}\frac{\mu^2-\nu^2}{\frac{1}{2}\mu^2})
\end{equation}
\noindent where $\varepsilon_\mu$ is the bit error rate of the
signal pulses, and $Q_\nu^L$ is the lower bound of the decoy pulse
transmittance, estimated conservatively as 10 standard deviations of
$Q_\nu$ from the measurement value, ensuring a confidence interval
of $1-1.5\times10^{-23}$. By assuming all bit errors are from single
photon pulses, the upper bound, $\varepsilon_1^U$, for the bit error
rate of the single photon pulses can be bounded as
\begin{equation}
\varepsilon_1^U=\frac{\varepsilon_\mu Q_\mu}{Q_1^L}
\end{equation}
\noindent With the bounds $Q_1^L$ and $\varepsilon_1^U$ available,
the secure bit rate is readily calculated, which will be shown
later.

We use a one-way fibre-optic QKD system with phase encoding,
schematically shown in Fig.~1. It uses two asymmetric Mach-Zender
interferometers for encoding and decoding. The sending (Alice) and
receiving (Bob) units are connected with a 25.3~km fibre spool. Both
the signal and decoy pulses are generated by a 1.55 $\mu$m pulsed
laser diode of fixed intensity operating at 7.143~MHz. The pulses
are modulated with an intensity modulator to produce the desired
ratio of signal pulse strength to decoy pulse strength, and are then
strongly attenuated to the desired level by an attenuator before
leaving Alice's apparatus. Synchronization is realized by
multiplexing a 1.3 $\mu$m clock signal with the quantum signals so
that clock and quantum signals can be sent through the same fibre.
An active stabilization technique \cite{yuan05} is used to ensure
continuous operation.

Two single photon detectors based on InGaAs avalanche photodiodes
are carefully prepared for the experiment. Care was taken to
minimize detector afterpulsing,\cite{ribordy98} which correlates
photon detection events of the signal and decoy pulses and prevents
faithful measurements of $Q_\mu$ and $Q_\nu$. The system detection
efficiency, including channel transmission loss (4.7~dB) and Bob's
loss (2.5~dB), was characterized at $1.95\times10^{-2}$. The sum of
the dark count rates for the two detectors was $9.4\times10^{-5}$
per pulse. Custom electronics were developed for driving the QKD
optics.

We implemented the one weak decoy state + BB84 protocol \cite{ma05}
in our experiment. The decoy pulse probability is chosen to be 1/4.
Each QKD session is set to be $1\times10^6$ sifted bits, including
both signal and decoy pulses, corresponding to approximately
$2\times10^6$ photon detection events. The signal and decoy pulse
intensities are chosen as $\mu=0.425$ and $\nu=0.204$ to provide
optimum secure bit rate.

We recorded the output from the decoy QKD experiment continuously
for four hours, during which a total of 372 QKD sessions were
executed, each lasting $38.7$~s on average. The measured bit error
rates ($\varepsilon_\mu, \varepsilon_\nu$) for signal and decoy
states, as shown in Fig.~2(a), are stable at 1.72$\%$ and 2.74$\%$
respectively. These values are consistent with  the measured
interferometer visibility, mis-compensation, and detector dark
counts. Transmittances $Q_\mu$ and $Q_\nu$, shown in Fig.~2(b),
fluctuate in tandem, and such fluctuation is attributed mainly to
variation of the channel transmission and the detection efficiency.
The sharp dips in $Q_\mu$ and $Q_\nu$ are due to polarization or
phase resets of the active alignment system, during which photons
were temporarily not counted. Figure 2(c) shows the measured ratio
of {$Q_\mu/Q_\nu$, which remains virtually constant despite the
larger fluctuations in $Q_\mu$ and $Q_\nu$. The measured ratio is
remarkably consistent with the theoretically expected (red line),
indicating the system has produced and detected both signal and
decoy pulses reliably.

The final secure bit rate, $R$, was obtained, after estimating
$Q_1^L$ and $\varepsilon_1^U$ using Eqns. (1) and (2),  by first
applying a modified Cascade\cite{brassard94} error correction
routine, followed by privacy amplification,\cite{bennett95} using
\begin{equation}
R\geq\{-N_\mu^{Sift}
f_{ec}H_2(\varepsilon_\mu)+N_1^{Sift}[1-H_2(\varepsilon_1^U)]\}/t
\end{equation}
\noindent where $N_\mu^{Sift}$ is the number of the sifted bits from
the signal pulses, $N_1^{Sift}=\frac{1}{2}Q_1^Lt$ the lower bound of
the number of the sifted bits for the single photon pulses, and $t$
is the QKD session time.
$H_2(\varepsilon)=-\varepsilon\log_2(\varepsilon)-(1-\varepsilon)\log_2(1-\varepsilon)$
is the binary Shannon entropy. $f_{ec}$ is the error correction
efficiency relative to Shannon limit. Our error correction module
gives $f_{ec}\approx1.10$ for the given bit error rate.

As shown in Fig.~\ref{fig:bitrate}, the secure bit rate remains
fairly stable and positive for all decoy sessions, with an average
value of 5.51~kbps. The rate is much higher than a non-decoy QKD
system. For a non-decoy system using the same detectors, we found
the maximum fibre length secure from the PNS attack to be just 9~km.
With quieter detectors, a secure bit rate of 43~bps was previously
achieved for 25~km of fiber.\cite{gobby04b} Thus, our implementation
of the decoy protocol has allowed more than two orders of magnitude
enhancement in the secure bit rate.

To detect the PNS and related attacks, Alice and Bob can monitor the
ratio $Q_\nu/Q_\mu$. In the current experiment, we expected a ratio
of $Q_\nu/Q_\mu\doteq\frac{\nu\eta+Y_0}{\mu\eta+Y_0}=0.486$, where
$\eta$ is the system detection efficiency and $Y_0$ is the dark
count rate. Significant deviation of the measured ratio from this
expected value indicates a PNS by Eve. In the PNS attack, Eve first
counts the number of photons in each pulse using a quantum
non-demolition measurement and preferentially transmits the
multiphoton pulses to Bob. Since the signal pulses are stronger than
the decoy pulses, and thus contain more multi-photon pulses, the PNS
will tend to reduce $Q_\nu/Q_\mu$. Figure~\ref{fig:pns} shows a
simulation, together with the experimental points, of the secure bit
rate dependence on the ratio $Q_\nu/Q_\mu$. The experimental data is
in excellent agreement with theory when the PNS is not present. With
the PNS attack present, the simulated secure bit rate decreases
rapidly with the ratio $Q_\nu/Q_\mu$. No secure bit rate is possible
when the ratio is merely $17\%$ smaller than expected, highlighting
the privacy amplification cost to remove information leaked by the
PNS attack. Such a sensitive dependence puts stringent requirements
on the preparation and measurement of the signal and decoy pulses.
The measured ratio may be slightly higher than expected due to
statistical uncertainty, or if Eve chooses to preferentially
transmit the single photon pulses. A more likely case of the
measured ratio exceeding the expected value is due to artifacts,
such as mis-calibration of signal and decoy intensity ratio and/or
detector afterpulsing. Such artifacts  would cause an
over-estimation of secure bit rate, thereby compromising security,
and must therefore be carefully eliminated. This has been achieved
in our experiment by careful calibration of the source and
detectors.

Secure bit rate is dependent on the session duration due to the
statistical certainty with which $\varepsilon_1^U$ and $Q_1^L$ can
be estimated. Longer sessions allow a more precise determinate and a
higher bit rate. It can be seen in Fig.~\ref{fig:size} that the
calculated secure bit rate increases with session duration,
approaching the maximum value for sessions exceeding 1000 seconds.
Due to statistical fluctuations, there is a minimum session time for
establishing a secure bit rate, found here to be around 0.4~s. We
perform decoy protocol QKD at three distinct session times: 9.2,
18.5, and 38.7~s, and results are shown also in Fig.~\ref{fig:size}.
Even with the shortest time of 9.2~s, a secure bit rate of 4.80~kbps
is achieved, which is about $78\%$ of the saturation rate with
infinite session time, demonstrating that long session times are not
necessary for decoy pulse QKD.

In summary, a practical one-way decoy QKD system has been
demonstrated over a 25.3~km fiber with an average bit rate of
5.51~kbps. It is unconditionally secure against all types of
attacks, including the PNS attack. We conclude that decoy pulses
improve the security and performance of weak pulse QKD. However,
sources and detectors must be calibrated accurately to avoid any
artifacts that may compromise security.

During preparation of this manuscript two other implementations of
decoy pulse QKD have appeared.\cite{peng06,rosenberg06} These show
operation over longer fibre spools but with much lower key
generation rates and are limited to local synchronization.

The authors gratefully acknowledge partial financial support from
the EU through the Integrated Project SECOQC.

\begin{figure}[t]
\begin{center}
\includegraphics[width=12cm]{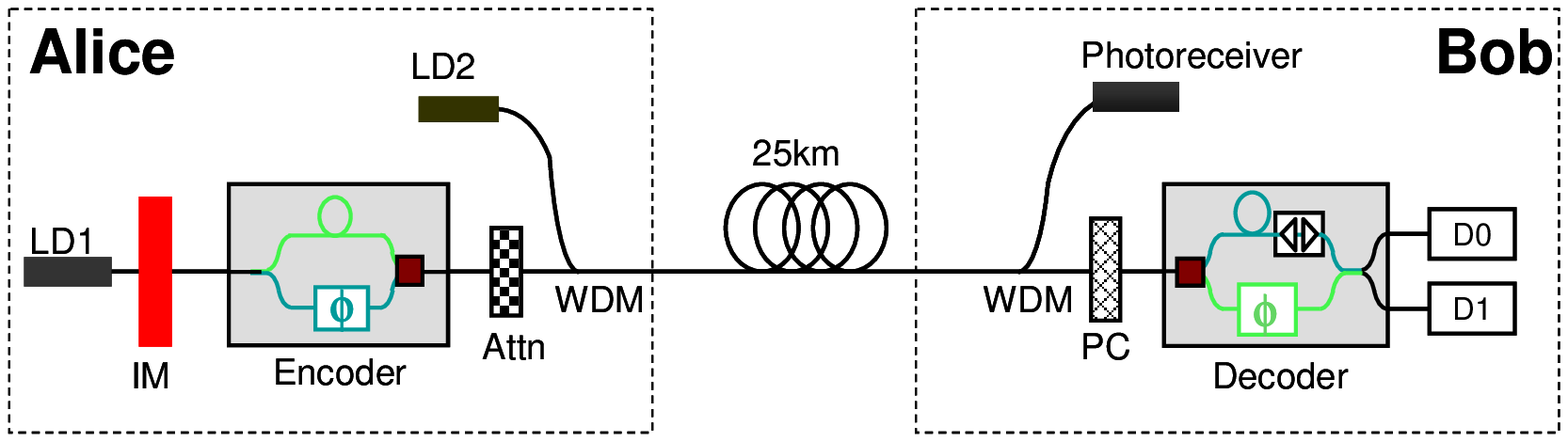}
\caption{
%\textbf{a}.
Schematics of optical layout of the one-way decoy quantum key
distribution system using BB84 phase-encoding; Attn: Attenuator; IM:
intensity modulator; LD1: signal laser diode at 1.55$\mu m$; LD2:
clock laser diode at 1.3$\mu m$; PC: polarisation controller; WDM:
wavelength division multiplexer; D0, D1: single photon detectors.}
\end{center}
\end{figure}

\begin{figure}[t]
\begin{center}
\includegraphics[width=10cm]{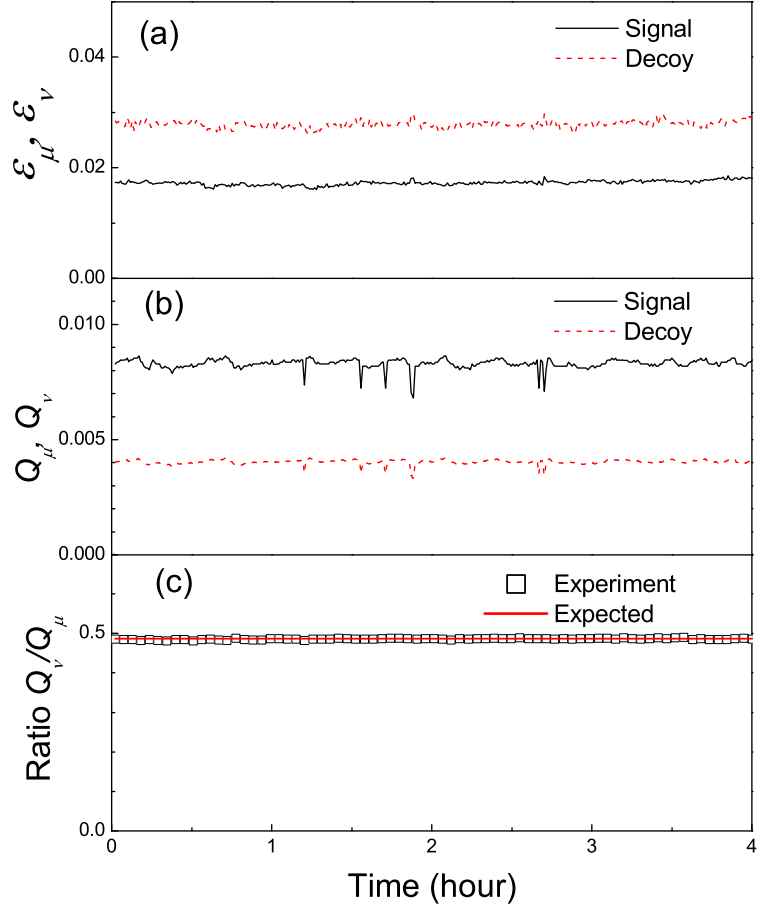}
\caption{Experimental results of a continuous series of decoy-QKD
sessions for four hours. (a) Bit error rates measured for the signal
and decoy pulses; (b) The measured transmittances for the signal
pulses ($Q_\mu$) and decoy pulses ($Q_\nu$); (c) The measured ratio
$Q_\mu/Q_\nu$ (empty squares) in comparison with the expected value
(line).}
\end{center}
\end{figure}

\begin{figure}[t]
\begin{center}
\includegraphics[width=10cm]{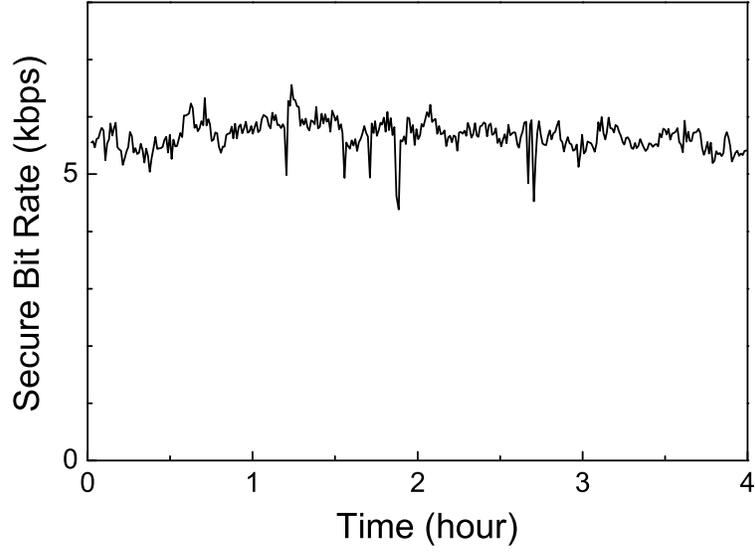}
\caption{\label{fig:bitrate}Secure bit rate as a function of time. }
\end{center}
\end{figure}

\begin{figure}[t]
\begin{center}
\includegraphics[width=10cm]{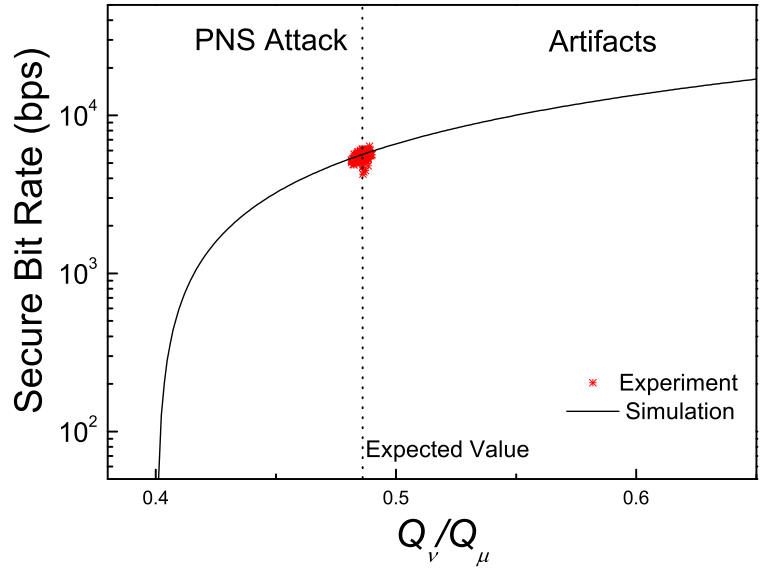}
\caption{\label{fig:pns}Simulated (solid line) and experimental
(symbols) secure bit rate dependence on the transmittance ratio
$Q_\nu/Q_\mu$. The vertical dotted line indicates the expected
ratio. }
\end{center}
\end{figure}

\begin{figure}[b]
\begin{center}
\includegraphics[width=10cm]{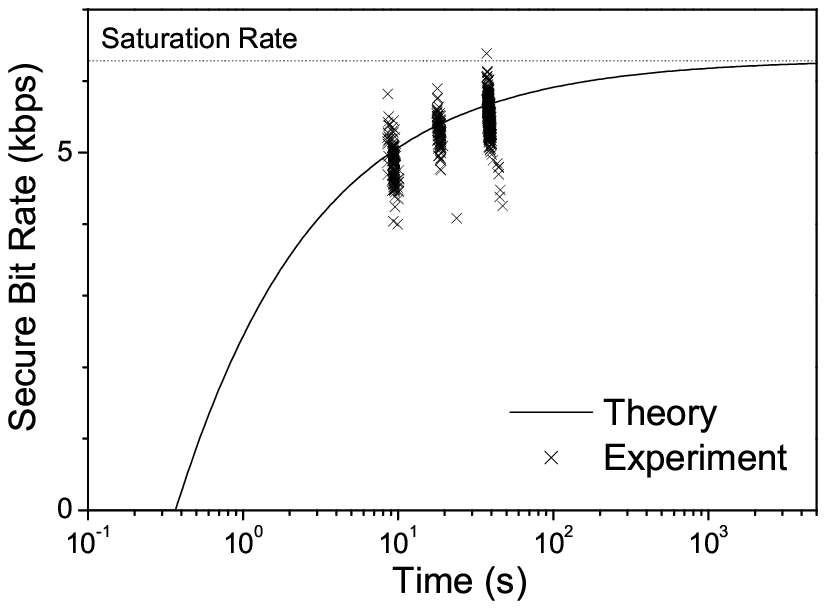}
\caption{\label{fig:size}Simulated (solid line) and experimental
(symbols) secure bit rate as a function of QKD session duration. The
dotted line shows the saturation secure bit rate with infinite QKD
sesssion time.}
\end{center}
\end{figure}


\begin{thebibliography} {99}
\bibitem{bennett84}
C. H. Bennett and G. Brassard, in Proc. of the IEEE Int. Conf. on
Computers, Systems and Signal Processing, Bangalore, India
(Institute of Electrical and Electronics Engineers, New York, 1984),
pp. 175-179.

\bibitem{gisin02} N. Gisin, G. Ribordy, W. Tittel, and H. Zbinden, Rev. Mod. Phys. {\bf 74}, 145-195(2002).

\bibitem{dusek06}
M. Dus\v{e}k, N. L\"utkenhaus, M. Hendrych, in \textit{Progress in
Optics}, Vol. \textbf{49}, edited by E. Wolf (Elsevier 2006).

\bibitem{mayers01}
D. Mayers, J. Asso. Comput. Machin. \textbf{48}, 351(2001).

\bibitem{shor00}
P. W. Shor and J. Preskill, Phys. Rev. Lett. \textbf{85}, 441(2000).

%\bibitem{ward05}
%M.~B.~Ward \textit{et al.},
%%O. Z. Karimov, D. C. Unitt, Z. L. Yuan, P. See, D. G. Gevaux, A. J. Shields, P. Atkinson, and D. A. Ritchie,
%Appl. Phys. Lett. \textbf{86}, 201111(2005).

\bibitem{townsend93}
P. D. Townsend, J. G. Rarity, and P. R. Tapster, Electron. Lett.
{\bf 29}, 634-635(1993).

\bibitem{gobby04a}
C. Gobby, Z. L. Yuan, and A. J. Shields, Appl. Phys. Lett. {\bf 84},
3762(2004).

\bibitem{brassard00}
G. Brassard et al., Phys. Rev. Lett. \textbf{85}, 1330(2000).

\bibitem{hwang03}
W. Y. Hwang, Phys. Rev. Lett. \textbf{91}, 050791(2003).

\bibitem{wang05}
X. B. Wang, Phys. Rev. Lett. \textbf{94}, 230503(2005).

\bibitem{lo05}
H-K. Lo, X. Ma and K Chen, Phys. Rev. Lett. \textbf{94},
230504(2005).

\bibitem{ma05}
X. Ma, B. Qi, Y. Zhao and H-K. Lo, Phys. Rev. A \textbf{72},
012306(2005).

\bibitem{zhao06}
Y. Zhao, B. Qi, X. Ma, H-K. Lo, and L. Qian, Phys. Rev. Lett.
\textbf{96}, 070502(2006); quant-ph/0601168.

\bibitem{inamori01}
H. Inamori, N. L\"utkenhaus, D. Mayers, quant-ph/0107017.

\bibitem{gottesman04}
D.~Gottesman, H.-K.~Lo, N.~L\"utkenhaus, and J.~Preskill, Quant.
Inf. Comput. \textbf{4}, 325(2004).

\bibitem{yuan05}
Z.~L.~Yuan and A.~J.~Shields, Opt. Express \textbf{13}, 660(2005).

\bibitem{ribordy98}
G. Ribordy, J. D. Gautier, H. Zbinden, and N. Gisin, Appl. Opt. {\bf
37}, 2272-2277(1998).

\bibitem{brassard94}
G. Brassard and L. Salvail, Lecture Notes Comp. Sci. \textbf{765},
410(1994).

\bibitem{bennett95}
C. H. Bennett, G. Brassard, C. Cr\'{e}peau, and U. M. Mauer, IEEE
Trans. Inf. Theory \textbf{41}, 1915(1995).

\bibitem{gobby04b}
C. Gobby, Z. L. Yuan, and A. J. Shields, Electron. Lett. {\bf 40},
1603-1605(2004).

\bibitem{peng06}
C. Z. Peng \textit{et al.},
%J. Zhang, D. Yang, W. B. Gao, H. X. Ma, H. Yin, H. P. Zeng, T. Yang, X. B. Wang, and J. W. Pan,
quant-ph/0607129.

\bibitem{rosenberg06}
D.~Rosenberg \textit{et al.},
%J.~W.~Harrington, P.~R.~Rice, P.~A.~Hiskett, C.~G.~Peterson, R.~J.~Hughes, J.~E.~Nordholt, A.~E.~Lita, and S.~W.~Nam,
quant-ph/0607186.

\end{thebibliography}
\end{document}